# A GPU-boosted high-performance multi-working condition joint analysis framework for predicting dynamics of textured axial piston pump

Xin Yao [a], Yang Liu, Jin Jiang[a], Yesen Chen, Zhilong Chen [a], Hongkang Dong [a], Xiaofeng Wei, Teng Zhang [a*], Dongyun Wang [a*]

[a] College of Engineering, Zhejiang Normal University, Jinhua, P.R.C., 321004

*Corresponding author: zhangteng@zjnu.edu.cn zsdwdy@zjnu.cn

**ABSTRACT:** Accurate simulation to dynamics of axial piston pump (APP) is essential for its design, manufacture and maintenance. However, limited by computation capacity of CPU device and traditional solvers, conventional iteration methods are inefficient in complicated case with textured surface requiring refined mesh, and could not handle simulation during multiple periods. To accelerate Picard iteration for predicting dynamics of APP, a **G**PU-boosted high-performance **M**ulti-working condition joint **A**nalysis **F**ramework (GMAF) is designed, which adopts Preconditioned Conjugate Gradient method (PCG) using Approximate Symmetric Successive Over-Relaxation preconditioner (ASSOR). GMAF abundantly utilizes GPU device via elevating computational intensity and expanding scale of massive parallel computation. Therefore, it possesses novel performance in analyzing dynamics of both smooth and textured APPs during multiple periods, as the establishment and solution to joint algebraic system for pressure field are accelerated magnificently, as well as numerical integral for force and moment due to oil flow. Compared with asynchronized convergence strategy pursuing local convergence, synchronized convergence strategy targeting global convergence is adopted in PCG solver for the joint system. Revealed by corresponding results, oil force in axial direction and moment in circumferential directly respond to input pressure, while other components evolve in sinusoidal patterns. Specifically, force and moment due to normal pressure instantly reach their steady state initially, while ones due to viscous shear stress evolve during periods. After simulating dynamics of APP and pressure distribution via GMAF, the promotion of pressure capacity and torsion resistance due to textured surface is revealed numerically, as several 'steps' exist in the pressure field corresponding to textures.

**KEY WORDS:** Dynamics of axial piston pump; GPU-boosted high-performance algorithm; Preconditioned Conjugate Gradient method; Approximate Symmetric Successive Over-Relaxation preconditioner; Synchronized convergence strategy

## 1. Introduction

Axial piston pump, possessing various industrial applications[1]-[6], is the key part competent of transmitting driven force in complicated working conditions with heavy load and high velocity. To promote its design, manufacture and maintenance, accurate simulation to dynamics of APP is of great essentiality. Significantly affected by the thin mineral oil between surfaces of chamber and piston, the dynamics of APP is predictable after obtaining the pressure field and lubrification effect due to oil flow governed by the Reynold's equation[7]-[10]. After determining the thickness field via the geometry equation[11]-[13], the algebraic system for pressure field is established via discretizing the Reynold's equation, in which FVM (finite volume method)[14][15] is the most popular method. Therefore, pressure field is constructed after solving the algebraic system and implementing boundary conditions. Commonly, establishment and solution to the algebraic system are executed on CPU[16]; however, even parallel computation is optional to enhance performance, limited by its capacity, CPU device could not analyze dynamics of APP in complicated cases with refined mesh (e.g., textured surface or surface roughness) during multiple periods at affordable cost. Given its novel capacity for massive parallel computation, GPU device is highly suggested to boost both establishment and solution sequences of the algebraic system[17][19][20].

Moreover, in current researches, iterative solvers are applied to solve the algebraic system, including Jacobian iteration, Gauss iteration, Successive symmetric over-relaxation method[10][18]; however, they are incapable of solving massive scale sparse algebraic system corresponding to increased number of DOFs in complicated case with refined meshes. Alternatively, PCG is an excellent choice[19]-[22], given its capability of handling sparse algebraic system and compatibility of massive-scale parallel computing. The preconditioner determines PCG's convergence and its rate[21]-[25], in which the Jacobian preconditioner, the main diagonal of matrix, is the most popular one, as it is considerably robust and grants convergence; however, it costs more iterations. Successive symmetric over-relaxation and incomplete Cholesky factorization preconditioner consume significantly less iterations[22]; however, their constructions and applications are not feasible for massive parallel computation, hindering further optimization on GPU device[23]. Alternatively, Approximate Symmetric Successive Over-Relaxation preconditioner (ASSOR) is an optimized choice, as it is naturally compatible with massive parallel computation[24][25]. Numerical examples in **Section 4.1** demonstrate that ASSOR could promote performance of PCG by 36% than Jacobian preconditioner, and GPU device could further boost its efficiency.

Utilizing Picard iteration for predicting dynamics of APP, Jacobian matrix incorporating partial derivatives of force and moment with respect to eccentricity and its rate are required and updated during sequential iterative step[12]. Applying first-order Euler difference method, at least 4 extra working conditions shall be analyzed for building each matrix, and such number increases to 8 for double Jacobian matrices[26][27]; thus, the establishment of Jacobian matrix is highly costive. All these extra conditions could be analyzed separately and sequentially, with each corresponding algebraic system built and solved via parallel computation. Such strategy is suitable for CPU device, as a compromise of its limited capacity of parallel computation. However, once mesh is refined in complicated case, CPU device fails to handle such arduous task at affordable cost[28]. Given the tremendous threads in GPU device, these multi-working conditions could be concurrently analyzed, with all sparse algebraic systems corresponding to different conditions established to forge a joint algebraic system. Following such principle, a **G**PU-boosted high-performance **M**ulti-working condition joint **A**nalysis **F**ramework (GMAF) is proposed in **Section 3.2**. Accompanied with PCG using ASSOR, GMAF jointly analyzes pressure distribution in different working conditions and conducts accordant numerical integrals for force and moment due to oil flow, so that establishment of Jacobian matrix is further boosted, as computational intensity is elevated and scale of parallel computation is expanded to maximize utilization of GPU device. GMAF enables and facilitates simulation to dynamics of APP in different periods at affordable cost, and it is also applicable in complicated cases with textured surface, heat convection, elastic deformation and so on. Results in **Section 4.3** and **4.4** reveals that GMAF further promotes computational efficiency by 124.72%~520.05%, compared with one executed via sequential GPU acceleration for each condition.

Utilizing GMAF to assemble and solve the joint algebraic system **Eq**.3.7, PCG solver at each Picard iterative step via shall be terminated after convergence, which is judged via either synchronized and asynchronized strategy. The first strategy **Eq**.3.9 reports convergence for entire joint algebraic system globally, while the second one **Eq**.3.10 finishes PCG solver after convergence of each subsystem is reached at local scale, so that PCG solver is completed after convergence of the last subsystem is met. The synchronized convergence strategy is conceptually direct and convenient to implement; however, theoretically, it requires more computation cost, since iteration for converged subsystem continues until global convergence of the entire system. Contrarily, the asynchronized strategy could save such excessive and redundant iterations: PCG iteration for any converged subsystem is stopped, the solver concentrates to the remaining subsystems until all of them are converged, namely, the scale of PCG solver's parallel SpMVs (Sparse matrix-vector multiplication) executed on GPU device is reduced. Realistically, given the fact that number of PCG iterations for each subsystem is rather close and adjustment to the scale of accordant kernel function performing SpMVs on GPU requires communication between CPU and GPU devices, moreover, the asynchronized strategy is stricter than the synchronized one, the asynchronized strategy demonstrates weaker performance than the synchronized strategy in **Section 4.2**. Thus, the synchronized strategy is chosen and utilized in PCG solver of GMAF.

After simulating dynamics of textured APP and accordant pressure distribution in entire time domain with multiple periods via GMAF, the resultants reveal that textures generate several smaller-amplitude 'steps' in pressure field (**Section 4.3** and **4.4**), which are the direct reason for promoted load capacity of textured piston. The presence of these steps is reasonable in both physical and mathematical aspects: physically, improvement of pressure capacity is only effective in region with textures, as oil film is much thicker in such region; thus, steps with same spatial positions as textures are generated, while pressure distributions in smooth areas are not affected. Judging from mathematical aspect, in textured regions with thicker oil film, the accordant coefficient in Reynold's equation increases, while gradient and rate of thickness is almost unchanged; thus, solution to the algebraic system leads to more smooth pressure distribution, as the pressure field possesses less significant gradient and amplitude. Only GMAF is capable of revealing such steps and pressure evolution due to textured piston in multiple periods, given its remarkable performance due to its advanced framework of parallel computation and utilization of PCG using ASSOR. Indicated by results in **Section 4.5**, the independently compiled codes of GMAF, which are irrelevant with any embedded functions in CUDA Toolkit, show its compatibility with different architectures of NVIDIA GPU devices. Its maximum performance is achieved on the state-of-art RTX 5000 series GPU device with Blackwell architectures, since massive threads, high-frequency processors, VRAM and other features of the GPU are abundantly utilized.

The reminder of the manuscript is organized as: the current researches about simulating dynamics of APP are briefly introduced in **Section 1**. Procedures for simulating dynamics of APP, PCG solver and details of conventional iteration method are listed in **Section 2**. Strategy of sequential GPU acceleration is presented in **Section 3**, as well as its defects and limitations. To overcome these limitations, GMAF using PGC with ASSOR is designed and presented in detail. In **Section 4**, several examples are presented for validation and comparison, and GMAF demonstrates its higher performance in simulation to dynamics of APP with/without textures. Finally, conclusions and future plans are in **Section 5**.

## 2. GPU-boosted simulation for dynamics of axial piston pump

The computational steps for predicting dynamics of APP are briefly discussed in this section, accompanied with induction of PCG and common strategy of GPU acceleration.

## 2.1 Establishment and compression of algebraic system for pressure field

To predict the lubrication effects of oil film, the Reynold's equation shall be solved, as it governs the pressure field of thin oil film between surfaces of piston and chamber [7]-[10]. Utilizing finite volume method (FVM), the Reynold's equation is transformed to the sparse algebraic system below[14][15]

$$\boldsymbol{Ap} = \boldsymbol{S} \quad (2.1)$$

where matrix $\boldsymbol{A}$ restores the coefficient according to spatial discretization, $\boldsymbol{p}$ and $\boldsymbol{S}$ the global vector of pressure and the source vector, respectively. Both coefficient matrix $\boldsymbol{A}$ (with size of $n \times n$) and source vector $\boldsymbol{S}$ are based on thickness field of oil film $h = h(\boldsymbol{e}, \dot{\boldsymbol{e}})$ determined by eccentricity $\boldsymbol{e}(e_1, e_2, e_3, e_4)$ and its rate $\dot{\boldsymbol{e}}(\dot{e}_1, \dot{e}_2, \dot{e}_3, \dot{e}_4)$,

$$\boldsymbol{A}_{[n \times n]} = \boldsymbol{A}[h(\boldsymbol{e}, \dot{\boldsymbol{e}})], \boldsymbol{S}_{[n \times 1]} = \boldsymbol{S}[h(\boldsymbol{e}, \dot{\boldsymbol{e}})] \quad (2.2)$$

since thickness field in textured piston is modelled by the geometric equation[11]-[13]

$$h(\boldsymbol{e}, \dot{\boldsymbol{e}}) = \sqrt{\left(R_c \cos\theta - \frac{e_3 - e_1}{L_F} y - e_1\right)^2 + \left(R_c \sin\theta - \frac{e_4 - e_2}{L_F} y - e_2\right)^2} - R_k + h_{Text} \quad (2.3)$$

where $h_{Text}$ represents the depth of texture ($h_{Text} = 0$ for smooth piston).

The specific expression of the algebraic system **Eq**.2.1 in each row is

$$A_P P_P + A_E P_E + A_W P_W + A_S P_S + A_N P_N = S_P \quad (2.4)$$

where subscripts represent the spatial discretization and the boundary conditions could be naturally incorporated via linear transform. For accurate simulation in complicated working conditions, especially in case with textured surface, the number of nodes (the size of matrix $n$), is recommended to 1~10 millions, which magnificently elevates the computational cost. Therefore, given the sparsity of coefficient matrix $\boldsymbol{A}$, compression is required to reduce cost of RAMs and computational efforts. A reasonable strategy compatible with parallel computation is to directly and separately build all diagonals of the matrix $\boldsymbol{A}$, as the coefficients are compressed and stored via

$$\boldsymbol{A}_P = \{A_{P_i}\}, i = 1,2, \cdots n, \boldsymbol{A}_S = \{A_{S_i}\}, i = 1,2, \cdots n - n_S, \boldsymbol{A}_N = \{A_{N_i}\}, i = 1,2, \cdots n - n_N \quad (2.5)$$

$$\boldsymbol{A}_E = \{A_{E_i}\}, i = 1,2, \cdots n - n_E, \boldsymbol{A}_W = \{A_{W_i}\}, i = 1,2, \cdots n - n_W \quad (2.6)$$

$$\boldsymbol{A}_{EB} = \{A_{E_i}\}, i = 1,2, \cdots n_E, \boldsymbol{A}_{WB} = \{A_{W_i}\}, i = 1,2, \cdots n_W \quad (2.7)$$

where $n_S$ and $n_N$ are the number of discretized points on southern and northern boundary (directly affected by input and output boundary conditions), $n_E$ and $n_W$ the eastern and western boundary, representing the continues boundary conditions, respectively. Following such strategy of compression, approximately $5n$ number of elements are restored and accessed and avoids all the unnecessary elements that will never be accessed, thus, the complexity of the coefficient matrix $\boldsymbol{A}$ and algebraic system is magnificently reduced.

## 2.2 Implementation of Preconditioned Conjugate Gradient

Given its natural compatibility of massive parallel computation, Preconditioned Conjugate Gradient is utilized to accelerates the solution procedure to the sparse algebraic system **Eq**.2.1, which is further boosted via GPU device. The Jacobian preconditioner, the main diagonal of the coefficient matrix $\boldsymbol{A}$, shares high adaptability and robustness,

$$\boldsymbol{M} = \boldsymbol{D} \quad (2.8)$$

where $\boldsymbol{A} = \boldsymbol{L} + \boldsymbol{D} + \boldsymbol{L}^T$ is the direct factorization of the coefficient matrix $\boldsymbol{A}$.

To efficiently utilize GPU device and further boost performance of PCG, the specific computational steps of standard PCG linear solver[19] is adjusted, as listed in **Table 1**, where the step 2, 3, 5 and 8 are modified (as highlighted in blue) to avoid repetitive calculation. According to **Section** 2.4, all SpMVs are executed via massive parallel computation on GPU, as different threads are designated to handle multiplication in each row of resultant vector simultaneously and separately without any mutual communication.

**Table 1** Adjusted PCG linear solver for algebraic system **Eq**.2.1

1. Let $\boldsymbol{p}_0$ be an arbitrary initial guess (zero vector is commonly set)
2. Initial settings $\boldsymbol{r}_0 = \boldsymbol{A}\boldsymbol{p}_0 - \boldsymbol{S}, \boldsymbol{z}_0 = \boldsymbol{M}^{-1}\boldsymbol{r}_0, \boldsymbol{u}_0 = \boldsymbol{z}_0, d_0 = \boldsymbol{r}_0^T \boldsymbol{z}_0$
3. Calculate intermediate vector $\boldsymbol{v} = \boldsymbol{A}\boldsymbol{u}_j$ and variable $\alpha_j = d_j/(\boldsymbol{u}_j^T \boldsymbol{v}), j = 0,1,2, \ldots, MaxIter$
4. Update solution $\boldsymbol{p}_{j+1} = \boldsymbol{p}_j + \alpha_j \boldsymbol{u}_j$
5. Calculate residual $\boldsymbol{r}_{j+1} = \boldsymbol{r}_j - \alpha_j \boldsymbol{v}$
6. If the convergence or stopping criterion is met, exit the loop
7. Update $\boldsymbol{z}_{j+1} = \boldsymbol{M}^{-1}\boldsymbol{r}_{j+1}$
8. Calculate intermediate variables $d_{j+1} = \boldsymbol{r}_{j+1}^T \boldsymbol{z}_{j+1}$ and $\beta_j = d_{j+1}/d_j$
9. Update searching vector $\boldsymbol{u}_{j+1} = \boldsymbol{z}_{j+1} + \beta_j \boldsymbol{u}_j$

The selection of preconditioner significantly affects performance of PCG solver. The previously mentioned Jacobian preconditioner **Eq**.2.8 is the most popular one, as it is considerably robust in various cases generating diagonally-dominant coefficient matrix; however, its performance and convergence rate are rather low, as more iteration is required (**Fig. 2**). Incomplete Cholesky factorization preconditioner $\boldsymbol{M} = \boldsymbol{L}^T \boldsymbol{L}$, an approximation of coefficient matrix $\boldsymbol{L}^T \boldsymbol{L} \approx \boldsymbol{A}$, costs significantly less iteration[22]; however, its establishment and implementation could not be conveniently booted via massive parallel computation. Successive symmetric over-relaxation preconditioner (SSOR) is[24]

$$M = (D + L)D^{-1}(D + L)^T \tag{2.9}$$

which avoids complicated factorization, as its establishment is direct; however, its implementation requires elimination to obtain inverses of triangular matrices, hindering parallel computation and its remarkable acceleration. Conclusively, excellent preconditioner shall not only possess rapid convergence rate and save iterations, but also be convenient and feasible to establish and implement via parallel computation, so that the performance of PCG solver is further promoted.

## 2.3 Picard iteration for predicting dynamics of APP

The equilibrium of APP requires that the general force $F$ of piston to be zero, namely,

$$F = F_E + F_I + F_O = 0 \tag{2.10}$$

where force $F_E$, $F_I$ and $F_O$ are the general force due to external load, inertia movement and oil flow, respectively. For convenience of programing and presentation, the Einstein Summation Convention is adopted and the general force is simplified to

$$F \equiv \{F_1, F_2, F_3, F_4\} \tag{2.11}$$

The evolution of eccentricity $e$ and its rate $\dot{e}$ shall maintain the equilibrium $F = 0$. After setting initial eccentricity $e^0$ and its rate $\dot{e}^0$, Picard iteration for predicting dynamics of APP proceeds.

During $k^{th}$ iteration for eccentricity and its rate at coming time $t_{l+1}$, the thickness field is $h[e^{(k)}, \dot{e}^{(k)}]$, based on which the algebraic system **Eq.**2.1 is accordingly built. After solving the system, the pressure field at $k^{th}$ iteration $p^{(k)}$ is obtained, as well as the general force $F^{(k)}$ and Jacobian matrices.

Utilizing the general iteration method[29], the equilibrium at next $(k + 1)^{th}$ iterative step is estimated using first-order Taylor's expansion,

$$F^{(k+1)} = 0 \approx F^{(k)} + \left.\frac{\partial F}{\partial e}\right|^{(k)} [e^{(k+1)} - e^{(k)}] + \left.\frac{\partial F}{\partial \dot{e}}\right|^{(k)} [\dot{e}^{(k+1)} - \dot{e}^{(k)}] \tag{2.12}$$

where the dynamics at $(k + 1)^{th}$ iterative step, $e^{(k+1)}$ and $\dot{e}^{(k+1)}$, shall more precisely fulfil the equilibrium.

Elements on two Jacobian matrices $F_{,e}$ and $F_{,\dot{e}}$ are established via numerical difference,

$$F_{,e} \equiv \partial F / \partial e \supset \left(\frac{\partial F}{\partial e}\right)^{(k)}_{ij} \equiv \left(\frac{\partial F_i}{\partial e_j}\right)^{(k)} \approx \frac{F_i[e_j^{(k)} + \Delta e_j] - F_i[e_j^{(k)}]}{\Delta e_j}, i,j = 1,2,3,4 \tag{2.13}$$

and

$$F_{,\dot{e}} \equiv \partial F / \partial \dot{e} \supset \left(\frac{\partial F}{\partial \dot{e}}\right)^{(k)}_{ij} \equiv \left(\frac{\partial F_i}{\partial \dot{e}_j}\right)^{(k)} \approx \frac{F_i[\dot{e}_j^{(k)} + \Delta \dot{e}_j] - F_i[\dot{e}_j^{(k)}]}{\Delta \dot{e}_j}, i,j = 1,2,3,4 \tag{2.14}$$

where components of general force $F_i[e_j^{(k)} + \Delta e_j]$ and $F_i[\dot{e}_j^{(k)} + \Delta \dot{e}_j]$ are derived from pressure field solely affected by increment $\Delta e_j$ and $\Delta \dot{e}_j$, $p^{(k)}[e_j^{(k)} + \Delta e_j, \dot{e}^{(k)}]$ and $p^{(k)}[e^{(k)}, \dot{e}_j^{(k)} + \Delta \dot{e}_j]$, respectively. Oher components of eccentricity and its rate remain unchanged and same as $e^{(k)}$ and $\dot{e}^{(k)}$, respectively.

Behind each element in the Jacobian matrices (**Eq.**2.13 and **Eq.**2.14) exists a sparse algebraic system for pressure field, namely,

$$\frac{\partial F_i}{\partial e_j} \leftarrow A(e_j + \Delta e_j)p(e_j + \Delta e_j) = S(e_j + \Delta e_j) \tag{2.15}$$

$$\frac{\partial F_i}{\partial \dot{e}_j} \leftarrow A(\dot{e}_j + \Delta \dot{e}_j)p(e_j + \Delta e_j) = S(e_j + \Delta e_j) \tag{2.16}$$

if first-order Euler difference is adopted.

For convenience, the accordant algebraic systems are simplified by number in subscript,

$$A_j p_j = S_j, j = 0,1,2,\dots,8 \tag{2.17}$$

where $A_0 p_0 = S_0$ is actually the system **Eq.**2.1 and other definitions are

$$A_j \equiv A(e_j + \Delta e_j), p_j \equiv p(e_j + \Delta e_j), S_j \equiv S(e_j + \Delta e_j), j = 1,2,3,4 \tag{2.18}$$

$$A_{j+4} \equiv A(\dot{e}_j + \Delta \dot{e}_j), p_{j+4} \equiv p(\dot{e}_j + \Delta \dot{e}_j), S_{j+4} \equiv S(\dot{e}_j + \Delta \dot{e}_j), j = 1,2,3,4 \tag{2.19}$$

Thus, numerical difference to build Jacobian matrix $F_{,e}$ (or $F_{,\dot{e}}$) via **Eq.**2.13 (or **Eq.**2.14) requires 4 extra difference pressure field $p^{(k)}[e_j^{(k)} + \Delta e_j, \dot{e}^{(k)}]$ affected to $\Delta e_j, j = 1,2,3,4$ (or $\Delta \dot{e}_j$), respectively, which magnificently increases computational cost. Considering about such computational cost and smaller amplitude of Jacobian matrix $F_{,e}$ in case with low velocity and pressure, the accordant second term on RHS of **Eq.**2.12 is commonly not considered and the iterative scheme in the conventional methods is simplified to[26][27]

$$F^{(k+1)} \approx F^{(k)} + \left.\frac{\partial F}{\partial \dot{e}}\right|^{(k)} [\dot{e}^{(k+1)} - \dot{e}^{(k)}] = 0 \tag{2.20}$$

Therefore, the rate $\dot{e}^{(k+1)}$ is

$$\dot{e}^{(k+1)} = \dot{e}^{(k)} - \left[\frac{\partial F}{\partial \dot{e}}\bigg|^{(k)}\right]^{-1} F^{(k)} \tag{2.21}$$

Applying the backward-difference scheme, the evolution of eccentricity is

$$e^{(k+1)} = e^{(k)} - \left[\frac{1}{\Delta t_l}\frac{\partial F}{\partial \dot{e}}\bigg|^{(k)}\right]^{-1} F^{(k)} \tag{2.22}$$

Such iteration repeats until dynamic convergence criteria considering residual of equilibrium is reached and the iterative calculation moves to next time point $t_{l+2}$.

In the conventional iterative scheme **Eqs**.2.21, the evolution of eccentricity $e$ is not directly concerned, despite the fact that it is indirectly updated via **Eq**.2.22 after obtaining rate $\dot{e}$ via **Eq**.2.21. Given the fact that the mathematical principle of total differential is violated, such simplified iteration scheme is unreliable, resulting in asynchronized evolution of eccentricity and rate, especially in complicated case with high pressure and velocity[29].

The general iteration method[29] utilizes double Jacobian matrices **Eq**.2.13 and **Eq**.2.14; however, limited by computation capacity of CPU device, such method could not be conducted at affordable cost. The method is executable on GPU device, especially when the establishment of double Jacobian matrices is conducted via GMAF.

## 2.4 Computational tasks executed on GPU

Aiming to predict dynamics of APP efficiently, massive parallel computation boosted via GPU device is a promising choice, given the multiple processors with tremendous threads, high speed VRAM, advanced architecture and other promising features of state-of-art GPU device. Different to the architecture of CPU device, GPU is majorly composited by ALUs (Arithmetic logic unit), which enables and facilitates large-scale parallelization.

Considering about the entire procedure for predicting dynamics of APP, there are certain several steps feasible of applying parallel computation on GPU, including:

I.) Parallel computing for building sparse algebraic system.

Following the compression strategy in **Section 2.1**, coefficients in the matrix $A$ are only stored in serval diagonals, and elements are actually separate; thus, different threads are accordingly designated to each discretized point, so that all the elements in the compressed matrix $A$ and vector $S$ are calculated and stored parallelly.

II.) Parallel computing for PCG solver.

Acting as the core of PCG solver, SpMVs are naturally compatible with parallel computation: the multiplication $Au_j$ in **Table 1**, referring to **Eq**.2.4, could be handled by separately applying $n$ threads for all $n$ rows simultaneously, so that the multiplication is accelerated. Besides, addition and inner product of vector could also be accelerated via similar parallel computation.

III.) Parallel computing for synthesizing general force due to oil flow.

The numerical integral for general force due to oil flow is conducted on each surface element confined by discretized points, thus, different threads according to these elements are separately designated to calculate synthetic force and moment based on pressure and viscous shear stress fields. Afterwards, the general force due to incremental component of eccentricity or its rate is obtained, so that the accordant Jacobian matrix is built and iteration for predicting dynamics of APP proceeds.

## 3. GPU-boosted multi-working condition joint analysis framework

The section incorporates modification to PCG and selection of ASSOR. To elevate utilization of GPU device, GPU-boosted high-performance multi-working condition joint analysis framework is designed in this section, which increases computational intensity and expands scale of parallel computation.

## 3.1 Approximate symmetric successive over-relaxation preconditioner

Based on SSOR **Eq**.2.9, one type of ASSOR is defined, whose inverse is approximately[24]

$$\{M^{-1}\}_{ii} \approx 1/\{(D+L)D^{-1}(D+L)^T\}_{ii} \tag{3.1}$$

The inverse of the main diagonal in SSOR is approximately regarded as the inverse of ASSOR, which facilitates implementation via parallel computation. However, without any relaxation factor, the performance of such ASSOR could not be further enhanced. Adding the relaxation factor, first type of ASSOR is defined

$$\{M^{-1}\}_{ii} \approx \omega(2-\omega)/\{(D+\omega L)D^{-1}(D+\omega L)^T\}_{ii} \tag{3.2}$$

Containing relaxation factor, second type of ASSOR and its approximated inverse are defined via[25]

$$M = KK^T = \frac{1}{\omega(2-\omega)}(D+\omega L)D^{-1}(D+\omega L)^T \tag{3.3}$$

and

$$M^{-1} \approx (2-\omega)\omega D^{-1}(I - \omega L^T D^{-1})D(I - \omega D^{-1}L)D^{-1} \tag{3.4}$$

respectively. Neumann polynomial expansion is applied to obtain the approximated inverse **Eq**.3.4.

As the approximate inverse **Eq**.3.4 is lengthy, calculation for $z_{k+1} = M^{-1}r_{k+1}$ (step 7 in **Table 1**) could be divided into two steps:

*First step*. Calculating intermediate variable

$$v = (I - \omega D^{-1}L)D^{-1}r_{k+1} \qquad (3.5)$$

*Second step*. Obtaining vector $z_{k+1}$

$$z_{k+1} = (2 - \omega)\omega(I - \omega D^{-1}L^T)v \qquad (3.6)$$

Such two-step modification avoids complicated multiplication of the approximated inverse, and facilitates its implementation with reduction of RAM cost.

Both first and second type of ASSOR are feasible of massive parallel computation, as their establishments and implementations are direct and avoid complicated and sequential calculation for inverses.

## 3.2 Multi-working condition joint analysis framework

As numerical difference for building Jacobian matrices in **Eq**.2.13 and **Eq**.2.14 are highly costive, a direct choice is to establish and solve the algebraic system **Eq**.2.17 via sequentially GPU acceleration (SGA). However, redundancy increases, as such sequential calculation for pressure field could not sufficiently maximize utilization of threads and VRAM in GPU. Alternatively, a GPU-boosted high-performance multi-working condition joint analysis framework is proposed, which simultaneously establish and solve the algebraic systems corresponding to all the increments, so that both Jacobian matrices are built via massive parallel computation.

Unlike sequential GPU acceleration for establishing and solving the algebraic system **Eq**.2.17, GMAF sets a joint algebraic system incorporating all 9 algebraic systems, namely,

$$\begin{bmatrix} A_0 & 0 & 0 & & & \\ 0 & A_1 & 0 & \cdots & & \\ 0 & 0 & A_2 & & & \\ \vdots & & & \ddots & & \vdots \\ & & & \cdots & A_7 & 0 \\ & & & & 0 & A_8 \end{bmatrix} \begin{bmatrix} p_0 \\ p_1 \\ p_2 \\ \vdots \\ p_7 \\ p_8 \end{bmatrix} = \begin{bmatrix} S_0 \\ S_1 \\ S_2 \\ \vdots \\ S_7 \\ S_8 \end{bmatrix} \Leftrightarrow A_G p_G = S_G \qquad (3.7)$$

where $A_G$, $p_G$ and $S_G$ are joint coefficient matrix, pressure vector and source vector, namely,

$$\{A_G\}_{i+j\times n} \equiv \{A_j\}_i, \{p_G\}_{i+j\times n} \equiv \{p_j\}_i, \{S_G\}_{i+j\times n} \equiv \{S_j\}_i, i = 0,1,2,\ldots.8, j = 1,2,\ldots.n^2 \qquad (3.8)$$

Therefore, compared with establishment and solution to the algebraic systems via SGA in **Section 2.4**, the scale of parallel computation in GMAF is increased by 8 times. As the scale increases, the computational intensity is elevated and GPU device could focus more on executing parallel computation rather than data communication or synchronization between threads, so that the efficiency is promoted.

Within GMAF, PCG using ASSOR is applied to solve the joint algebraic system **Eq**.3.7. After the joint pressure vector incorporating all 9 working conditions is obtained, the synthetic force and moment of oil film in all working conditions are simultaneously calculated via parallel computation as well. Thus, the double Jacobian matrices are built and iteration for predicting dynamics of piston is conducted afterwards.

For simplicity, the major steps of GMAF for predicting dynamics of APP are listed below:
I.) Establish joint algebraic system via massive parallel computation, accompanied with boundary conditions.
II.) Solve the joint algebraic system via PCG using ASSOR, update and rebuild pressure field.
III.) Conduct numerical integral for synthetic force and moment of oil film via parallel computation.
IV.) Build Jacobian matrices using numerical difference based on the synthetic force and moment.
V.) Perform iteration for predicting dynamics of APP and repeat above steps.

Given the structure of the joint algebraic system **Eq**.3.7 and computational steps of GMAF, such framework is also applicable of analyzing elastic deformation and heat conduction in piston and chamber, as well as heat convection of oil film. Moreover, GMAF could handle thermo-elastic-hydrodynamic coupling analysis efficiently.

## 3.3 Synchronized and asynchronized convergence strategies

Two strategies determining convergence of the joint algebraic system **Eq**.3.7 are optional.

*Synchronized convergence strategy*: the strategy pursuits convergence of joint algebraic system,

$$\frac{\left\| A_G p_G^{(k+1)} - S_G \right\|_2}{\|S_G\|_2} \leq \epsilon_{PCG} \qquad (3.9)$$

where $\epsilon_{PCG}$ is the criterion judging convergence.

The strategy seeks for global convergence and finishes iteration after solution to the entire joint algebraic system is obtained. The strategy is direct and convenient to implement, and the size of parallel SpMVs on GPU device requires no further adjustment. However, it conducts PCG iterations on converged subsystems until global convergence, which requires excessive and unnecessary costs.

*Asynchronized convergence strategy*: it seeks for convergence of subsystems **Eq**.2.17 in 9 working conditions,

$$\frac{\left\| A_j p_j^{(k+1)} - S_j \right\|_2}{\|S_j\|_2} \leq \epsilon_{PCG}, j = 1,2,\ldots 9 \qquad (3.10)$$

The asynchronized strategy finishes PCG iteration to each subsystem locally and it adjusts size of accordant kernel functions executing SpMVs, vector addition and inner product, so that excessive or redundant iterative calculation is saved theoretically.

The synchronized convergence strategy **Eq**.3.9 is weaker than the asynchronized one **Eq**.3.10, as the error is averaged and smoothed by the entire joint algebraic system with 9 times multiplied DOFs. The synchronized strategy could reduce number of PCG iterations, while the accuracy may be affected consequently, which could be granted via constrained convergence criterion $\epsilon_{PCG}$.

## 4. Examples for validation and comparison

In this section, performance of PCG solver via serial computation on CPU and parallel computation on GPU is compared, as well as effects of different preconditioners on performance of PCG are analyzed. Moreover, the magnificent acceleration achieved via GMAF in cases evolved with smooth and textured piston is demonstrated.

### 4.1 Performance of PCG on CPU/GPU

The detailed parameters and initial settings of APP are listed in **Fig. 8** and **Table 8** of **Appendix A**. On uniformly-distributed mesh, the accordant algebraic system for pressure is established and solved. The codes are executed on desktop equipped with CPU AMD® Ryzen 7600X and GPU NVIDIA® RTX 5080. The performance of executing PCG with Jacobian preconditioner via serial computation on CPU and parallel computation on GPU in cases with different meshes are discussed and compared in **Fig. 1** (convergence criterion of PCG is $10^{-6}$).

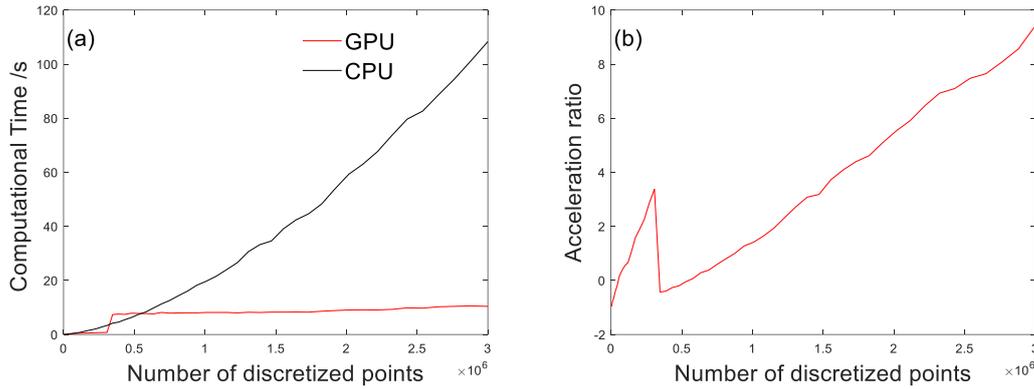

**Fig. 1** Performance of PCG on different meshes. (a) Computational Time on CPU/GPU. (b) Acceleration ratio.

As presented by black curve in **Fig. 1**(a), computational time of PCG solver via serial computation on CPU is approximately proportional to number of discretized points; however, red curve indicates that computational time via parallel computation on GPU almost remains unchanged, even if the scale of algebraic system expands apparently. Such phenomenon is accordant to two devices' different capacity of executing parallel computation, as the CPU device, possessing limited number of cores, could handle much fewer threads simultaneously. Contrarily, the GPU device shares massive CUDA units and stream processors, accompanied with higher-frequency VRAM and other features, is highly capable of multi-threads processing. The acceleration ratio amplifies once refined mesh is utilized, since CPU device meets its limit of parallel computation while stream processors of the GPU device is not abundantly utilized, even if the number of discretized points exceeds million.

To achieve strict convergence of $10^{-12}$ in case with mesh size of $2000 \times 1600$, PCG solver using Jacobian preconditioner **Eq**.2.8, SSOR **Eq**.2.9 and ASSOR I **Eq**.3.2 and ASSOR II **Eq**.3.4 (relaxation factor $\omega = 1.8$) are applied, whose results are presented in **Fig. 2**(a). The effect of relaxation factor $\omega$ on performance of PCG using ASSOR II is analyzed (convergence criterion is $10^{-6}$), as a series of relaxation factors are set and tested, $\omega = 0.18i + 0.1, i = 1,2, \dots ,10$. The number of iterations via PCG with ASSOR II versus different relaxation factor $\omega$ is listed in **Fig. 2** (b), demonstrating that relaxation factor around $1.4{\sim}1.6$ is of optimized performance and could save iterations.

Demonstrated by the curves, PCG using ASSOR II **Eq**.3.4 converges more rapidly after merely 3738 iterations. The performance of PCG solver using Jacobian preconditioner, SSOR and ASSOR I are quite similar, as they share similar patterns of residuals and require 6352, 6352 and 6519 iterations, respectively. However, the computational times corresponding to SSOR, ASSOR I, ASSOR II and Jacobian preconditioners are of significant difference, as they consume $1680.10s$, $215.21s$, $226.57s$ and $246.41s$, respectively. Such apparently different computational efforts are reasonable, since implementation of SSOR evolves elimination for inverses of matrices, while more SpMVs are required during applying ASSOR II, even its number of iterations are the minimum among all methods.

Thus, in the manuscript, PCG using ASSOR II is utilized and abbreviation ASSOR represents ASSOR II **Eq**.3.4. Both PCG solvers using Jacobian preconditioner and ASSOR are accelerated via parallel computation on GPU device, whose resultants are listed in **Table 2** (convergence criterion is $10^{-6}$).

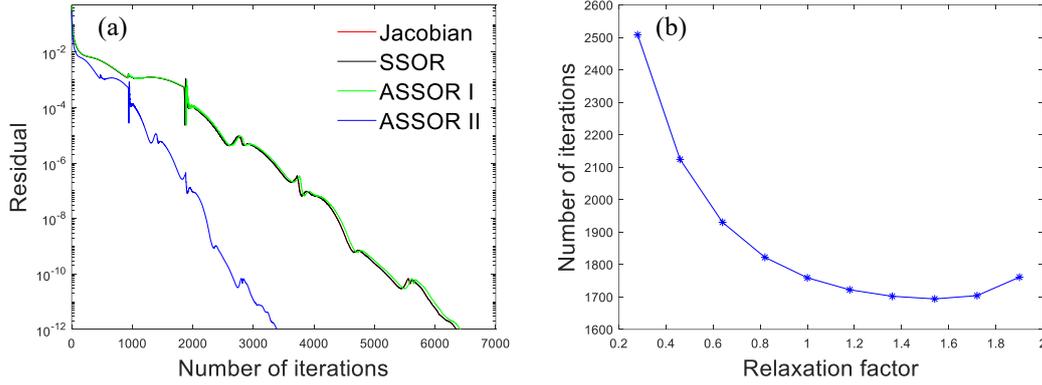

**Fig. 2** Performance of PCG on mesh size of $2000 \times 1600$. (a) Residual of PCG versus number of iterations. (b) Number of iterations via PCG with ASSOR II versus relaxation factor.

**Table 2** Computational cost of PCG on mesh size $2000 \times 1600$

|  | Number of iterations | Time cost / second | Time cost per iteration / second |
| --- | --- | --- | --- |
| PCG (Jacobian) | 3326 | 120.774 (CPU) | 0.0363 |
|  | 3308 | 5.860 (GPU) | $1.7715 \times 10^{-3}$ |
| PCG (ASSOR) | 1719 | 117.518 (CPU) | 0.0684 |
| $\omega = 1.8$ | 1709 | 3.747 (GPU) | $2.1925 \times 10^{-3}$ |

Due to precision of GPU and CPU device, numbers of iterations are of slight difference; however, the GPU device could boost the speed of codes using the Jacobian preconditioner and ASSOR by magnificent margin of 1960.99% and 3036.32%, respectively, which are majorly due to boosted calculations for SpMVs, vector addition and inner product via massive parallel computation on the state-of-art GPU device. Precisely, ASSOR shares 56.39% higher efficiency on GPU device than the Jacobian preconditioner.

### 4.2 Comparison of synchronized and asynchronized convergence strategies

Actual performances of the synchronized and asynchronized convergence strategy are tested. For pistons with short and long textures located at the bottom areas (displayed in **Fig. 10**), the computational costs of solving the joint algebraic system **Eq**.3.7 via two strategies are listed in **Table 3** (relaxation factor $\omega = 1.6$), where 'AVG number of PCG iterations' is the resultant of total number of PCG iterations via the asynchronized strategy averaged over all 9 working conditions.

**Table 3** Computational cost of GMAF on textured piston (mesh: $2000 \times 1600$)

|  |  | Number of PGC iterations | Time cost / second | AVG time per iteration / second |
| --- | --- | --- | --- | --- |
| Short textured piston |  |  |  |  |
| Synchronized | PCG Jacobian | 6921 | 61.9315 | $8.9483 \times 10^{-3}$ |
|  | PCG ASSOR | 3583 | 50.3125 | $1.4042 \times 10^{-2}$ |
| Asynchronized | PCG Jacobian | 6921 | 81.9129 | $1.1835 \times 10^{-2}$ |
|  | PCG ASSOR | 8028 | 128.9036 | $1.6057 \times 10^{-2}$ |
| Long textured piston |  |  |  |  |
| Synchronized | PCG Jacobian | 9784 | 88.4215 | $9.0374 \times 10^{-3}$ |
|  | PCG ASSOR | 5236 | 73.7309 | $1.4082 \times 10^{-2}$ |
| Asynchronized | PCG Jacobian | 9784 | 116.4602 | $1.1903 \times 10^{-2}$ |
|  | PCG ASSOR | 10001 | 130.3075 | $1.3029 \times 10^{-2}$ |

Comparing computational times via PCG using ASSOR in **Table 2** and **Table 3**, it is concluded that solution to discretized Reynold's equation for pressure distribution on textured surface requires more PCG iterations than one on smooth surface. Therefore, simulation for dynamics of textured piston requires excessive computational efforts, which could not be efficiently handled by conventional iterative solver on CPU device, and only via GMAF could such simulation be conducted at affordable cost.

The computational times in both cases indicate that the synchronized convergence strategy **Eq**.3.9 consumes less time actually. The asynchronized strategy **Eq**.3.10 could theoretically save excessive and redundant efforts via selectively concentrating PCG iteration on remaining subsystems beyond convergence; however, such strategy requires communication between CPU and GPU devices: the size of kernel function handling SpMVs on GPU device shall be adjusted after convergence of any subsystem is achieved. Even on the state-of-art RTX 5080 GPU devices with broader band width and quicker VRAM, such communication for adjustment is costive and hinders consistent parallel computation, as specifically demonstrated by the results in **Table 3**: two strategies consume same iterations, while average time per iteration is increased by 32.264% and 31.710% in short and long textured cases via the asynchronized strategy, respectively. Moreover, as theoretically deduced in **Section 3.3** and demonstrated in **Table 3**, the asynchronized strategy, stricter than the synchronized one, seeks convergence of each subsystem locally and performs more iterations until convergence, namely, during implementation of PCG using ASSOR, the asynchronized strategy requires 124.058% and 91.005% more PCG iterations than the ones via the synchronized strategy in cases with short and long textures, respectively. Therefore, the asynchronized convergence strategy may not be efficient in realistic simulation, and the synchronized convergence strategy is preferred during implementation of GMAF.

## 4.3 Performance of GMAF on smooth piston

GMAF is applied to simulate pressure distribution in 9 working conditions simultaneously, as well as sequential GPU acceleration (SGA). Their corresponding performances of assembling and solving algebraic systems are listed in **Table 4**, where the number of PCG iterations via SGA is averaged over 9 working conditions (the synchronized convergence strategy is adopted and ASSOR is applied with relaxation factor $\omega = 1.8$).

**Table 4** Computational cost of GMAF and SGA on smooth piston

| | | Number of PCG iterations | Time cost / second | AVG time per iteration / second | Acceleration with respect to SGA |
|---|---|---|---|---|---|
| mesh $400 \times 360$ | GMAF Assembly | | $1.510 \times 10^{-4}$ | | 399.01% |
| | PCG Jacobian | 777 | 0.7477 | $9.6229 \times 10^{-4}$ | 520.05% |
| | PCG ASSOR | 421 | 0.5297 | $1.2582 \times 10^{-3}$ | 429.47% |
| | SGA Assembly | | $7.535 \times 10^{-4}$ | | |
| | PCG Jacobian | 778 | 4.6361 | $6.6211 \times 10^{-4}$ | |
| | PCG ASSOR | 421 | 2.8046 | $7.4020 \times 10^{-4}$ | |
| mesh $800 \times 760$ | GMAF Assembly | | $1.501 \times 10^{-4}$ | | 401.20% |
| | PCG Jacobian | 1546 | 3.3941 | $2.1954 \times 10^{-3}$ | 182.97% |
| | PCG ASSOR | 853 | 2.7932 | $3.2746 \times 10^{-3}$ | 124.72% |
| | SGA Assembly | | $7.523 \times 10^{-4}$ | | |
| | PCG Jacobian | 1548 | 9.6044 | $6.8938 \times 10^{-4}$ | |
| | PCG ASSOR | 853 | 6.2768 | $8.1761 \times 10^{-4}$ | |

Indicated by results in **Table 4**, the computational cost of assembling the algebraic systems is not affected by number of meshes, as the accordant computational times in two meshes are rather close. The number of PCG iterations in each working condition via GMAF and SGA are almost the same. GMAF expands the size of parallel computation and elevates computational intensity, so that the capacity of GPU for massive parallel computation is utilized abundantly, as the GPU device consumes approximately 195W and its load remains in high percentage of 95~99% during execution. Therefore, the performances in **Table 4** indicate that the GMAF could not only accelerate assembly of algebraic system by approximately 400%, but also boost the solution procedure by 124.72%~520.05%. Precisely, the computational cost of executing PCG via GMAF is approximately proportional to number of iterations. However, such linear relation is not satisfied for PCG via SGA, as additional cost is required for communication between CPU and GPU device, generating redundant time and idle gap between sequential simulations to different working conditions. Moreover, utilizing SGA, the size of parallel computation is limited, hindering abundant utilization of GPU's full capability: the GPU device consumes approximately 103W and utilization of the device suddenly drops to almost zero during redundant time between simulation to different working conditions.

Comparing time cost in two cases with different mesh settings, the efforts to solve the joint system via GMAF is almost linearly proportional to the number of meshes: the number is increased by 4.222 times on two meshes, and the computational times of PCG solvers using Jacobian preconditioner and ASSOR are increased by 4.539 and 5.273 times, respectively. However, as mesh is refined, the averaged times for PCG using ASSOR are increased by more apparent margins than ones via PCG using Jacobian preconditioner and become considerably larger, namely, the averaged computational cost during each iteration of PCG solver using ASSOR is more sensitive to the number of meshes.

Affected by the time-varying input pressure acting on bottom of piston (**Fig. 9**) and constant output pressure $0.5 Mpa$ on the top of piston, on mesh with size of $2000 \times 1600$, the dynamics of smooth APP (detailed settings in **Fig. 8** and **Table 8**) during total 10 periods with $360 \times 10$ time steps are presented in **Fig. 3**, where the general iteration method[29] is utilized. Conducting GMAF, the computational time is $89653.5s$, namely, the average time cost per time step is $24.90s$, which is consumed during joint analysis of totally 9 working conditions for establishing double Jacobian matrices and 4~6 Picard iterations for approximately achieving equilibrium **Eq**.2.12 at each time step, demonstrates the novel performance via GMAF.

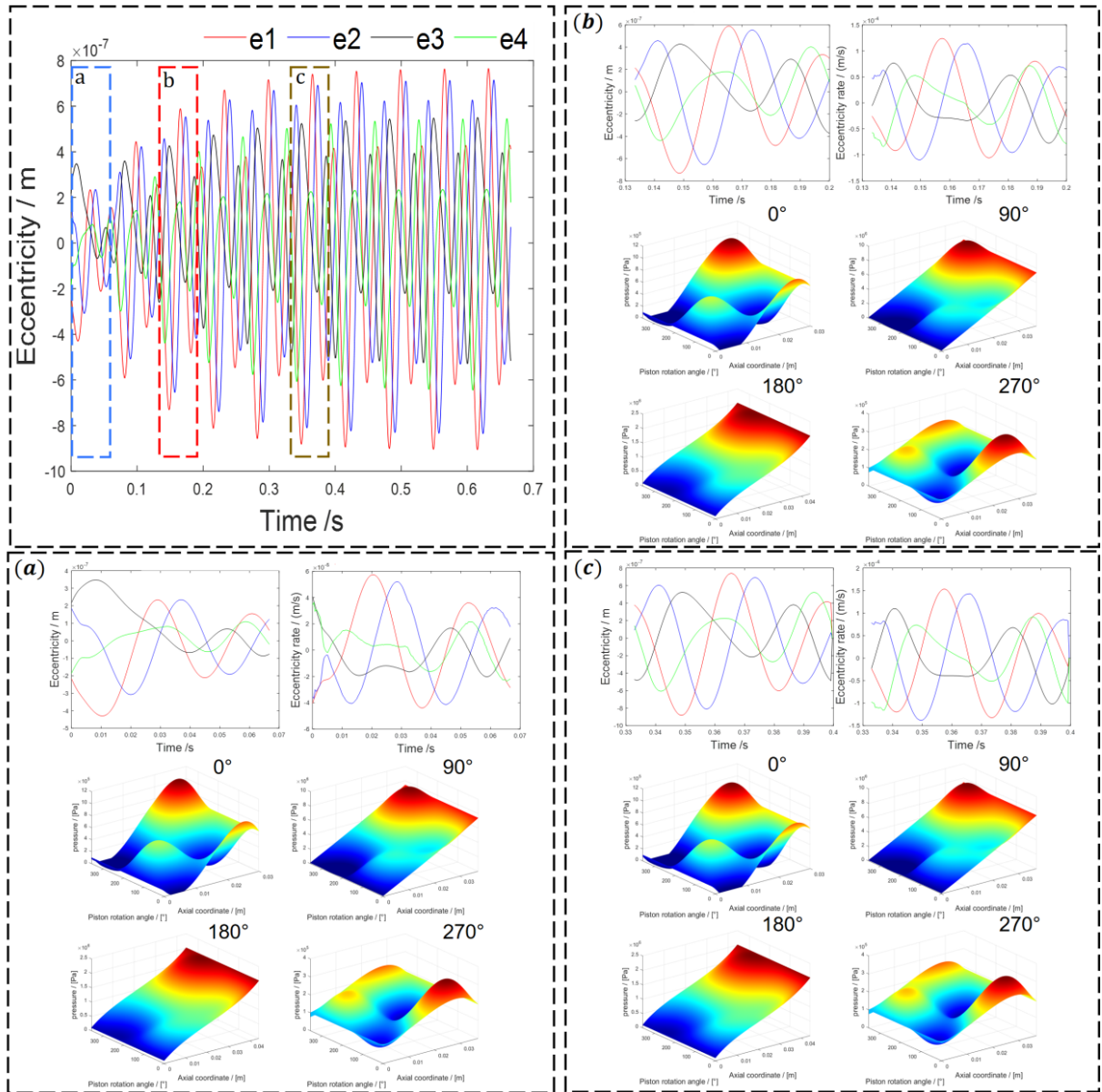

**Fig. 3** Time history of eccentricity and its rate, pressure distribution in different periods and angles in smooth APP. (a)~(c): results in the 1st, 3rd and 6th periods.

The resultant via GMAF reasonably models the evolution of eccentric and pressure field. During initial stage, apparent oscillation occurs, as shown in **Fig. 3**(a), where rate of eccentricity is more rapid and pressure distribution possesses more significant peaks. After several periods, the amplitude of eccentricity rate decreases and becomes more stable, resulting in more smooth pressure distribution, as shown in **Fig. 3**(b). Eventually, the evolution of eccentricity and pressure distribution are both in stable patterns, as the dynamics of APP reaches its steady state, as shown in **Fig. 3**(c).

The force and moment of oil flow in smooth piston are presented in **Fig. 4**, where red and black, blue and green curves represent resultants in circumferential and axial directions, respectively. Subfigures (a)~(d) indicate that the force and moment due to normal pressure in different periods are of negligible differences, namely, the force and moment reach their steady state rapidly at the initial period. However, the force in axial direction and moment in circumferential direction due to normal pressure both appropriately respond to input pressure in **Fig. 9**, while the force in circumferential direction and moment in axial direction evolve in sinusoidal patterns. Contrarily, the force and moment due to viscous shear stress are different, as their patterns actually evolve and approach the steady state after several periods. The force and moment in circumferential and axial directions are of different amplitudes and phase lags are detected, namely, the force and moment in circumferential direction reach their peaks before the ones in axial meets their peaks. Moreover, the moment in axial direction shows more oscillations, responding to the time-varying input.

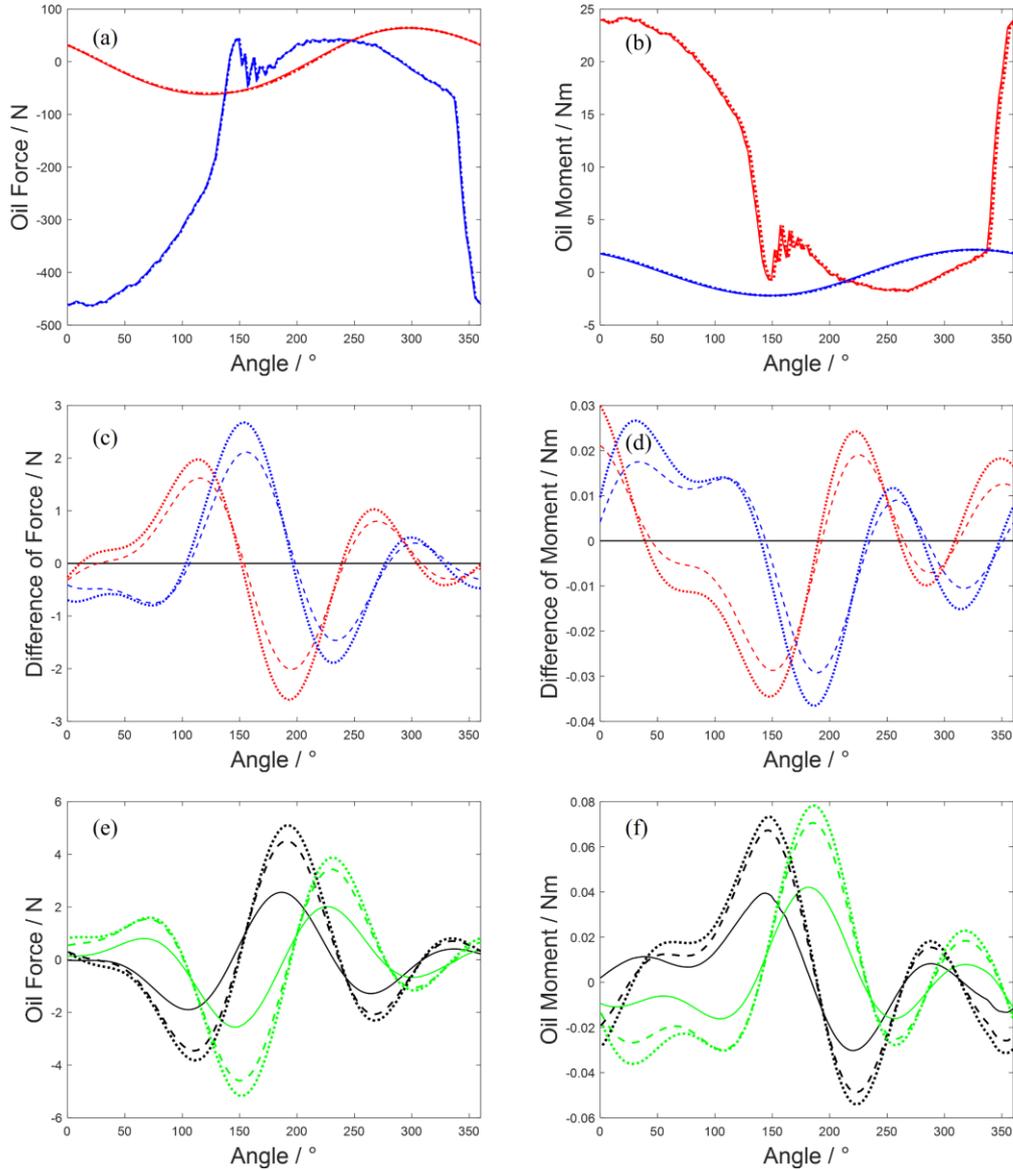

**Fig. 4** Force and moment due to oil flow in smooth piston. (a)(b) Oil force and moment due to normal pressure. (c)(d) Difference of force and moment with respect to results in 1$^{st}$ period. (e)(f) Oil force and moment due to viscous shear stress (solid, dashed line and dotted line represent result in 1$^{st}$, 3$^{rd}$ and 9$^{th}$ periods, respectively).

## 4.4 Performance of GMAF on textured piston

For pistons with short and long textures located at the bottom areas (**Fig. 10**), GMAF is applied to analyze the joint algebraic system **Eq**.3.7, as well as sequential GPU acceleration. Their corresponding performances are listed in **Table 5** and **Table 6**, respectively (relaxation factor $\omega = 1.6$ is set for ASSOR).

Even in complicated cases with textures, **Table 5** and **Table 6** demonstrate the higher performance of GMAF in both establishment and solution to the joint algebraic system. The speeds for establishment are promoted via GMAF by great margins of 278.06~330.13% and 262.40~268.22% in cases with short and long textures, respectively. On meshes with same size, the averaged computational times per each PCG iteration are of approximately 20~30% difference, as implementation of ASSOR requires excessive SpMVs, vector additions, inner products and other calculation, even the costs for these specific calculations are majorly determined by size of algebraic system. However, more PCG iterations are conducted in cases with textures, specifically, in cases with long texture, the number of iterations further increases, comparing **Table 5** to **Table 6**, respectively. Therefore, the total computational time for simulating dynamics of textured piston elevates. The computational time for case with long textures via GMAF is $142170.1s$ and the average time cost per time step is $39.49s$, and such increase of computational cost is accordant to the results in **Table 4** and **Table 5**.

**Table 5** Computational cost of GMAF and SGA on piston with short textures

| | | Number of PCG iterations | Time cost / second | AVG time per iteration / second | Acceleration with respect to SGA |
|---|---|---|---|---|---|
| mesh $400 \times 360$ | GMAF Assembly | | $2.001 \times 10^{-4}$ | | 430.13% |
| | PCG Jacobian | 1446 | 1.4016 | $9.6929 \times 10^{-4}$ | 515.67% |
| | PCG ASSOR | 757 | 0.9417 | $1.2440 \times 10^{-3}$ | 416.25% |
| | SGA Assembly | | $8.607 \times 10^{-4}$ | | |
| | PCG Jacobian | 13023 | 8.6293 | $6.6262 \times 10^{-4}$ | |
| | PCG ASSOR | 6813 | 4.8615 | $7.1356 \times 10^{-4}$ | |
| mesh $800 \times 760$ | GMAF Assembly | | $2.311 \times 10^{-4}$ | | 278.06% |
| | PCG Jacobian | 3151 | 7.2556 | $2.3026 \times 10^{-3}$ | 181.38% |
| | PCG ASSOR | 1631 | 5.2627 | $3.2267 \times 10^{-3}$ | 124.72% |
| | SGA Assembly | | $8.737 \times 10^{-4}$ | | |
| | PCG Jacobian | 28368 | 20.4155 | $7.1967 \times 10^{-4}$ | |
| | PCG ASSOR | 14679 | 11.7009 | $7.9712 \times 10^{-4}$ | |

**Table 6** Computational cost of GMAF and SGA on piston with long textures

| | | Number of PCG iterations | Time cost / second | AVG time per iteration / second | Acceleration with respect to SGA |
|---|---|---|---|---|---|
| mesh $400 \times 360$ | GMAF Assembly | | $2.382 \times 10^{-4}$ | | 268.22% |
| | PCG Jacobian | 1989 | 1.9817 | $9.9633 \times 10^{-4}$ | 493.51% |
| | PCG ASSOR | 1048 | 1.2842 | $1.2254 \times 10^{-3}$ | 416.08% |
| | SGA Assembly | | $8.771 \times 10^{-4}$ | | |
| | PCG Jacobian | 17910 | 11.7616 | $6.5671 \times 10^{-4}$ | |
| | PCG ASSOR | 9432 | 6.6275 | $7.0266 \times 10^{-4}$ | |
| mesh $800 \times 760$ | GMAF Assembly | | $2.431 \times 10^{-4}$ | | 262.40% |
| | PCG Jacobian | 4365 | 9.9644 | $2.2828 \times 10^{-3}$ | 179.51% |
| | PCG ASSOR | 2320 | 7.5686 | $3.2623 \times 10^{-3}$ | 123.08% |
| | SGA Assembly | | $8.810 \times 10^{-4}$ | | |
| | PCG Jacobian | 39294 | 27.8510 | $7.0879 \times 10^{-4}$ | |
| | PCG ASSOR | 20880 | 16.8843 | $8.0864 \times 10^{-4}$ | |

Similar to setting in **Section 4.3**, the dynamics of APP with short texture during total 10 periods are presented in **Fig. 5**. The general patterns of dynamics are similar to ones of smooth piston, namely, after initial oscillation in **Fig. 5**(a), the eccentricity and its rate gradually become stable in **Fig. 5**(b) and **Fig. 5**(c). However, due to short textures, the evolution of eccentricity is amplified by margin of 10~15%, accompanied with acceleration of evolution of eccentricity rate. The first component of eccentricity vector, $e_1$, becomes more active and its amplitude exceeds other components due to short textures. The amplified eccentricity and accelerated rate due to textures lead to more smooth evolution patterns, since the sudden drops and subtle oscillations in **Fig. 3** are avoided and alleviated in **Fig. 5**. The peaks of pressure field are reduced, resulting in more uniform distribution.

The force and moment due to oil flow in piston with short textures are presented in **Fig. 6**, where solid, dashed line and dotted line represent result in 1st, 3rd and 9th periods, respectively. Similar to results in **Fig. 4**, the force in axial direction and moment in circumferential direction due to normal pressure both respond to input pressure, while the rest evolve in sinusoidal patterns. The evolution of force and moment due to normal pressure reach their corresponding steady state rapidly at the initial stage, while the ones due to viscous shear stress present apparent evolution patterns. Unlike the curves in **Fig. 4**, the force in axial direction and moment in circumferential are specifically different, as the axial force starts from approximately $-100N$ and rapidly drops to its negative maximum around $-490N$, which is of slightly larger amplitude than one in **Fig. 4**(a), $-470N$, namely, the piston with short textures is of higher capacity responding to external load. Precisely, at approximately 150°, the oscillation of axial force due to normal pressure is alleviated in the textured piston, while the smooth one yield evolution pattern with more fluctuations. At the end of period, the axial force returns to $-100N$ as the initial result in **Fig. 6**(a), while in the smooth piston, the axial force suddenly drops to its minimum in **Fig. 4**(a). The drop of axial force is larger in smooth piston than one in the textured piston, which could amplify the noise and oscillation, namely, the performance and service life of textured piston is higher than the smooth one.

Comparison between subfigures (e)(f) of **Fig. 4** and **Fig. 6** indicates that the forces due to viscous shear stress are of less difference. Specifically, due to textures, the amplitudes of oil force are reduced; however, the accordant axial force remains smooth pattern with higher amplitude encountering input pressure, which provides more robust and continuous axial support. Due to textures, the axial moment also possesses more stable and higher patterns corresponding to external load, namely, the piston could resist intensified torsion during implementation of external load, even is amplitude is slightly reduced. Conclusively, owing to GMAF, the promotion of axial support and torsion resistance due to textures is revealed, as well as evolutionary patterns of dynamics of APP and pressure distribution. Moreover, the promoted supports due to textures also smooth dynamics of piston and alleviate oscillation, reducing noise and expanding service life.

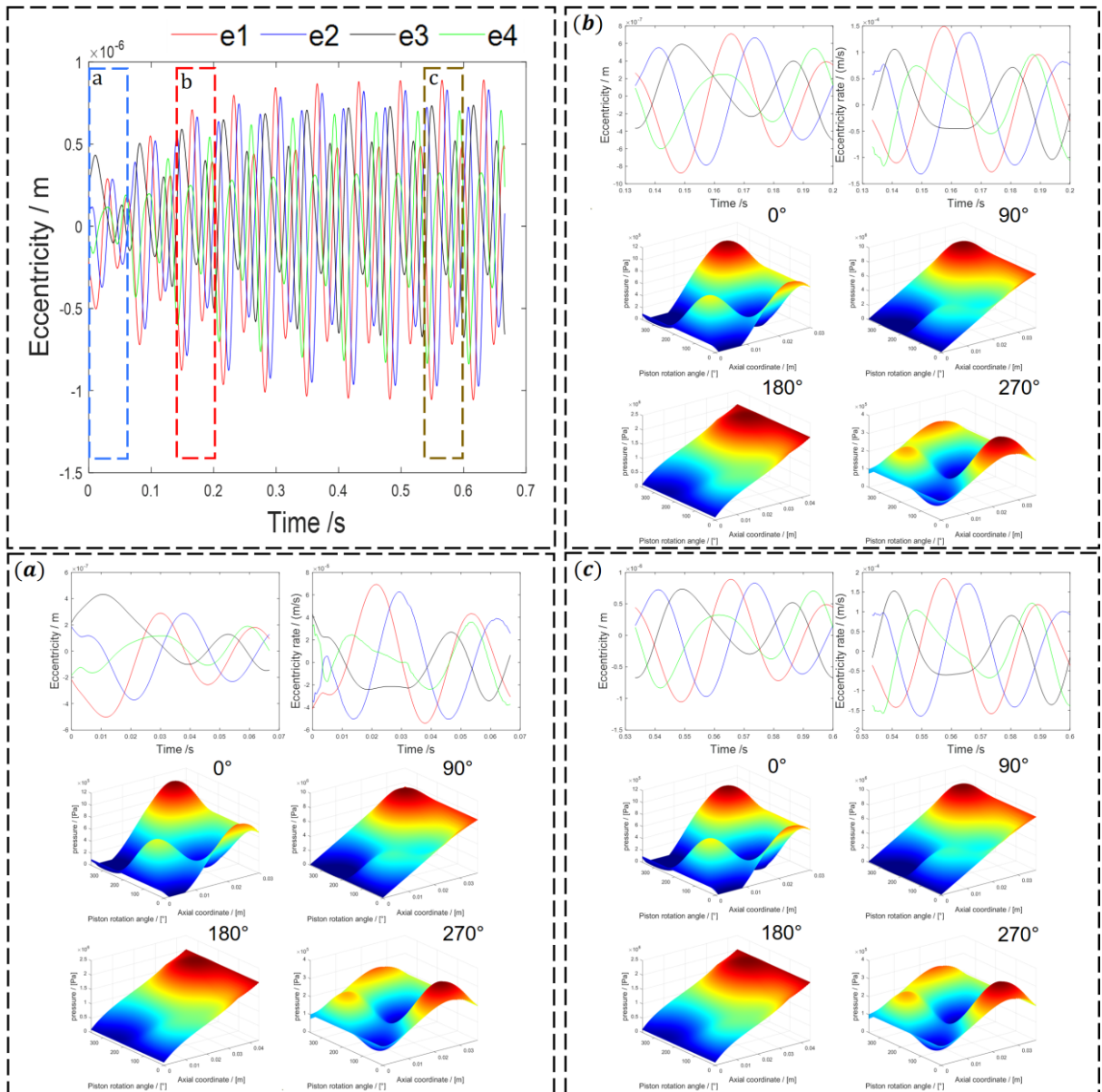

**Fig. 5** Time history of eccentricity and its rate, pressure distribution in different periods and angles in APP with short texture. (a)~(c): results in the 1$^{st}$, 3$^{rd}$ and 9$^{th}$ periods.

To demonstrate the specific effect of textures, the pressure distributions on smooth and short texture piston are presented in **Fig. 7**. Despite the general patterns of pressure are quite similar, the pressure distributions near bottom of piston are affected by textures, as little dots exist in **Fig. 7**(b). Precisely, the magnified subfigures reveal that several steps are uniformly distributed in pressure field, which are collocated to the textures in **Fig. 10**. Under same pressure input and outlet, the pressure field affected by textures and these steps are of smaller amplitude, compared with the pressure in smooth piston. Conclusively, **Fig. 7** demonstrates that the textured piston is of higher-pressure capacity.

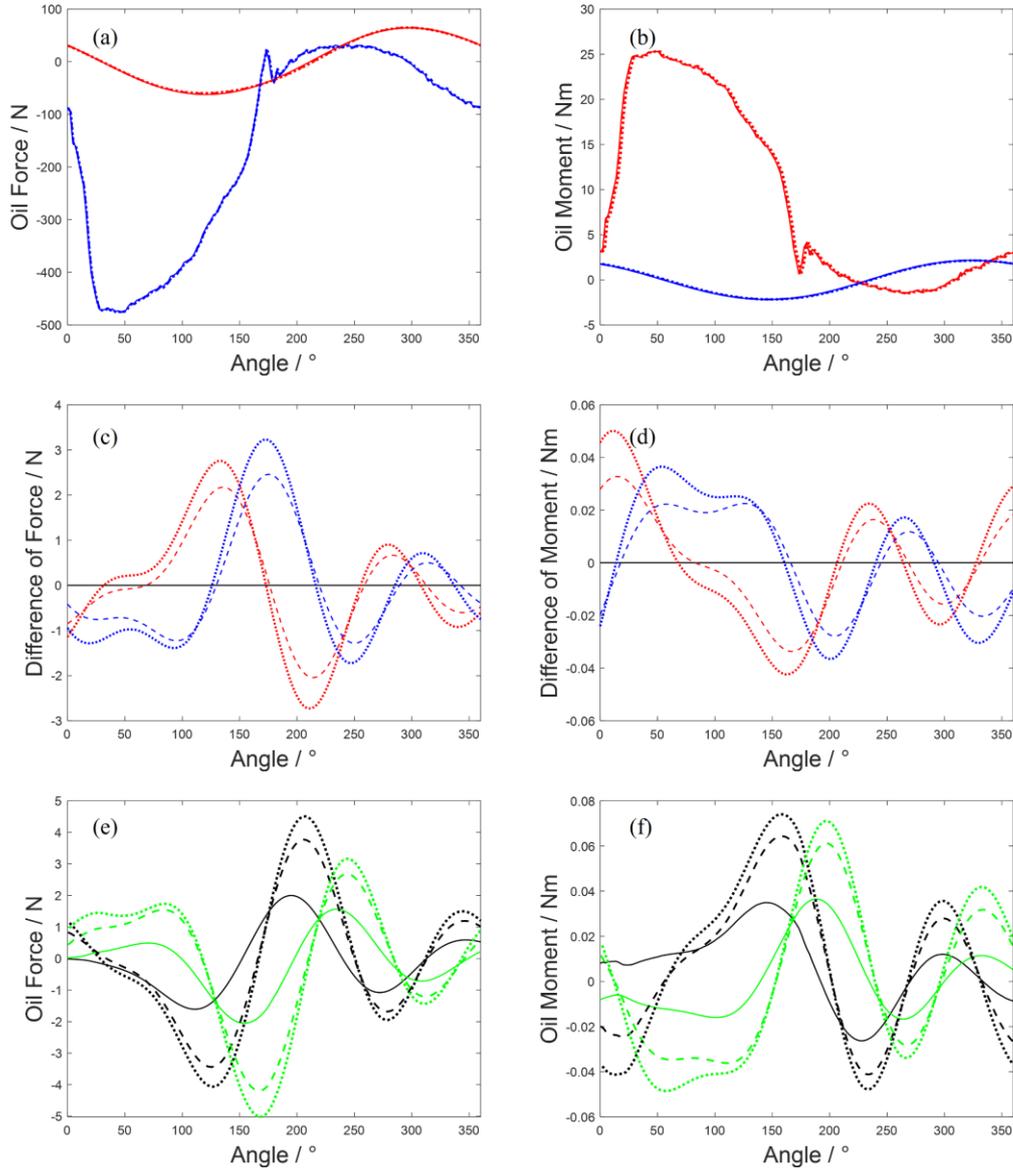

**Fig. 6** Force and moment due to oil flow in piston with short textures. (a)(b) Oil force and moment due to normal pressure. (c)(d) Difference of force and moment with respect to force in 1$^{st}$ period. (e)(f) Oil force and moment due to viscous shear stress.

The smaller-amplitude pressure field and several steps could be explained from both physical and mathematical aspects. Given the depth of texture $h_{Text} = 20\mu m$, the oil film is much thicker in the textured region, which could bear more significant pressure physically. As the improvement of pressure capacity is only effective in region with textures, steps with same spatial locations as textures are generated, while pressure distributions in rest areas are not affected. Judging from mathematical aspect, in textured regions, the accordant elements of coefficient matrix $A$ in the discretized algebraic system **Eq**.2.1 increases, while the source vector $S$ is almost unchanged; thus, pressure field with less significant gradient and amplitude is solved, resulting in more smooth pressure distribution.

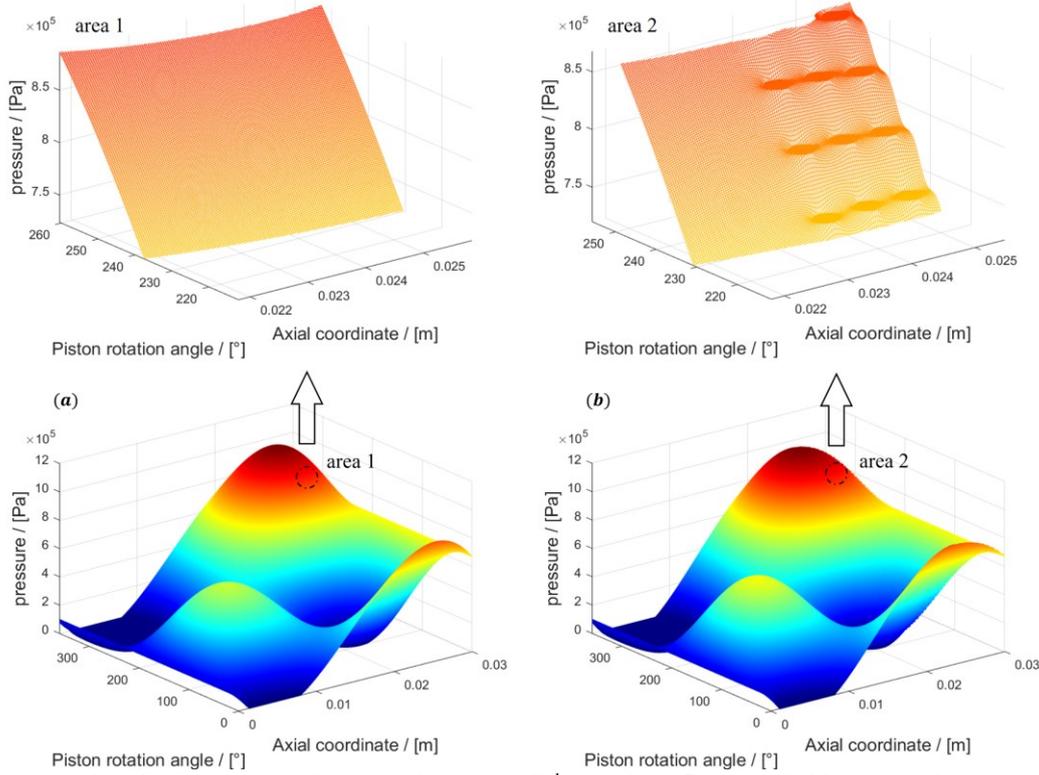

**Fig. 7** Pressure distribution on smooth/textured piston in 3$^{rd}$ period. (a) Pressure field on smooth piston. (b) Pressure field on textured piston.

## 4.5 Comparison of performance of GMAF on different GPUs

The independently compiled codes of GMAF for simulating dynamics of APP are executed on desktops equipped with NVIDIA® RTX 4090D and RTX 5080 separately, so that the determining factors for performance of GMAF are analyzed. Referring to **Table 4**, **Table 5** and **Table 6** computational times on RTX 4090D and RTX 5080 in corresponding cases are listed in **Table 7**.

**Table 7** Computational cost of GMAF and SGA on smooth piston (mesh: $800 \times 760$)

|  |  |  | Time cost on RTX 5080 / second | Time cost on RTX 4090D / second | Reduction |
|---|---|---|---|---|---|
| smooth piston | GMAF | PCG Jacobian | 3.3941 | 3.0905 | 8.945% |
|  |  | PCG ASSOR | 2.7932 | 2.3040 | 17.514% |
|  | SGA | PCG Jacobian | 9.6044 | 10.5406 | −9.748% |
|  |  | PCG ASSOR | 6.2768 | 5.9435 | 5.310% |
| short textured piston | GMAF | PCG Jacobian | 7.2556 | 6.2379 | 14.026% |
|  |  | PCG ASSOR | 5.2627 | 4.4941 | 14.605% |
|  | SGA | PCG Jacobian | 20.4155 | 21.3393 | −4.525% |
|  |  | PCG ASSOR | 11.7009 | 11.8043 | −0.884% |
| long textured piston | GMAF | PCG Jacobian | 9.9644 | 8.6938 | 12.751% |
|  |  | PCG ASSOR | 7.5686 | 6.3792 | 15.715% |
|  | SGA | PCG Jacobian | 27.8510 | 29.7632 | −6.866% |
|  |  | PCG ASSOR | 16.8843 | 17.0020 | −0.697% |

Indicated by results in **Table 7**, RTX 4090D more efficiently performs GMAF than RTX 5080, as the averaged reduction of computational times are approximately 13.926% and such reduction is more apparent in cases with textures, especially when ASSOR is utilized. However, RTX 4090D is of weaker performance in executing SGA for PCG, even both the power consumptions of two devices are around $220 \sim 240W$ in all cases.

Such phenomena are could be explained by the specific characters of two GPU devices, as listed in **Table 9** of **Appendix A**, where both devices are manufactured by $5nm$ technology. Since the power consumptions on both devices do not reach their maximum TDPs, the TDPs are not listed. Indicated by **Table 9**, even RTX 4090D device shares apparently better performance for FP32, its performance for FP16 is much lower; thus, neither these performances of two devices could directly and simply explain the actual difference of their computational times in **Table 7**. Given the apparently fewer CUDA cores and transistors (RTX 5080 possesses 26.32% fewer CUDA cores and 40.24% fewer transistors, while its GPU clock and VRAM clock are 4.44% and 42.59% higher than ones of RTX 4090D, respectively), RTX 5080 still achieves comparable efficiency, thus, the state-of-art Blackwell architecture promotes its actual performance, even it is of lower grade than RTX 4090D. If similar grade devices are compared, RTX 5090D will certainly achieve magnificently higher performance. Conclusively, the independently complied codes of GMAF could abundantly utilize the advanced characters of GPU device with the Blackwell architecture, e.g., higher frequency GPU clock, VRAM clock and so on.

## 5. Conclusion

Based on results, discussions and comparisons in this manuscript, serval conclusions are made:

*First.* Preconditioned conjugate gradient method is selected and modified to solve the algebraic system for pressure field based on the Reynold's equation. GPU device could boost the speed of codes using PCG with Jacobian preconditioner and ASSOR by magnificent margin of 1960.99% and 3036.32%, respectively, compared with the Jacobian iteration method on CPU. The acceleration ratio increases in cases with refined mesh and more discretized points. After evaluating the performance of different preconditioners, ASSOR is selected, which shares 56.39% higher efficiency than Jacobian preconditioner and SSOR.

*Second.* To accelerate Picard iteration for predicting dynamics of axial piston pump, a GPU-boosted high-performance multi-working condition joint analysis framework is designed, which increases computational intensity and expands the scale of massive parallel computation, as it not only builds and solves the joint algebraic system, but also conducts numerical integral for synthetic force and moment due to oil flow in different working conditions simultaneously. The independently compiled codes of GMAF could abundantly utilizes the GPU device with Blackwell architecture, as it accelerates establishment of algebraic system by approximately 400%, and boosts the solution procedure by 124.72%~441.73%, compared with ones via sequential GPU acceleration.

*Third.* Utilizing the synchronized convergence strategy, GMAF could efficiently predict reliable dynamics of both smooth and textured piston. Revealed by comparison, the textured surface generates several 'steps' in pressure distribution and promotes the pressure capacity of textured piston, which is theoretically reasonable at both mathematical and physical aspects. After applying GMAF, results of both smooth and textured piston during multiple periods are obtained, indicating that the force and moment due to normal pressure reach their steady state instantly during first period, while the ones determined by viscous shear stress actually evolve and reach their stable patterns after several periods.

*Fourth.* GMAF reveals that affected by oil flow, the force in axial direction and moment in circumferential direction due to normal pressure directly respond to the input pressure, as well as the force and moment in circumferential direction due to viscous shear stress; however, the rest forces and moments are less affected by the input and evolve in sinusoidal patterns. Given its high performance and adaptability, GMAF not only enable and facilitate the simulation to dynamics of both smooth and texture pistons in multiple periods at affordable cost, but is also feasible in complicated case incorporating heat convection and elastic deformation.

**Funding**. This work was supported by the National Natural Science Foundation of China (D. Wang grant No. 52375068, X. Wei grant No. 12202399).

**Appendix A** Settings of axial piston pump and input pressures and characters of GPU devices

**Fig. 8** and **Table 8** include detailed geometric parameters and initial settings of APP.

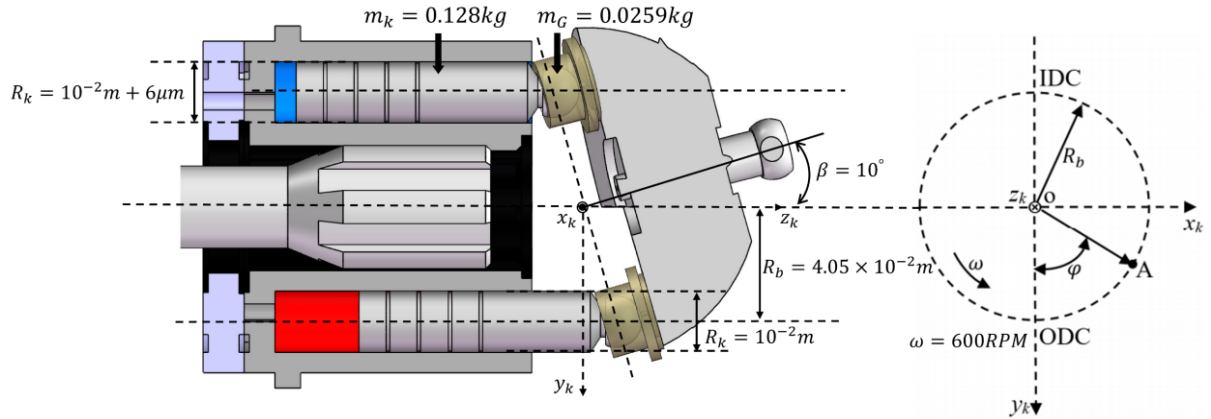

**Fig. 8** Geometric parameters of axial piston pump

**Table 8** Detailed parameters and initial settings of APP

| | | | |
|---|---|---|---|
| Radius of axial piston $R_k$ | $10^{-2}m$ | Radius of cylinder plunger $R_c$ | $10^{-2}m + 6\mu m$ |
| Radius of cylinder $R_b$ | $4.05 \times 10^{-2}m$ | Minimum coupling length $L_{Fmin}$ | $3 \times 10^{-2}m$ |
| Swashplate inclination angle | $10°$ | Revolutions per minute | $600 RPM$ |
| Mass of piston $m_K$ | $0.128 kg$ | Mass of slipper $m_G$ | $0.0259 kg$ |
| Eccentricity $e$ | | $(-0.2, 0.2, 0.2, -0.2) \times 10^{-6} m$ | |
| Rate of eccentricity $\dot{e}$ | | $(-3.78, 3.78, 3.78, -3.78) \times 10^{-7} m/s$ | |
| $\Delta e$ for numerical difference | $10^{-9}m$ | $\Delta \dot{e}$ for numerical difference | $10^{-8} m/s$ |

**Fig. 9** and **Fig. 10** present the time-varying input pressure and spatial distributions of textures on piston.

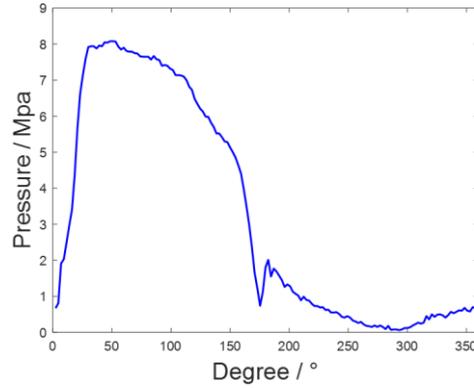

**Fig. 9** Time-varying input pressures measured from real-time experiments in **Section 4.3** and **4.4**.

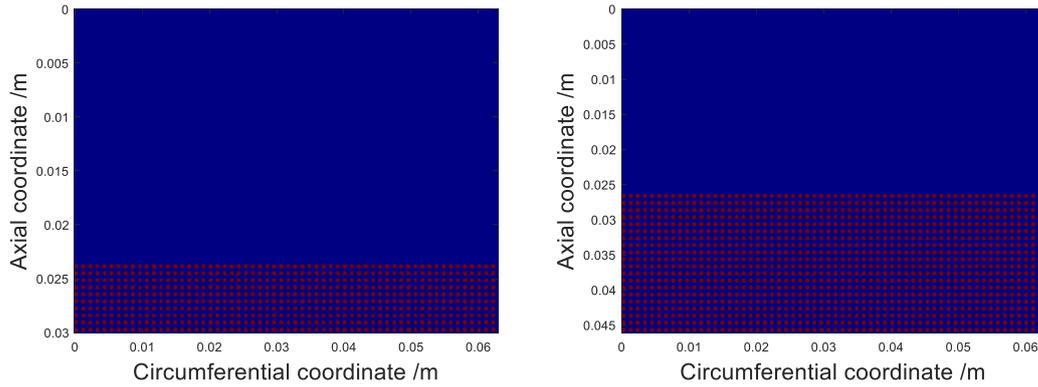

**Fig. 10** Texture distribution on piston (depth of texture $h_{Text} = 20\mu m$). LEFT: $60 \times 10$ Short texture. RIGHT: $60 \times 20$ Long texture.

**Table 9** Characters of RTX 5080 and RTX 4090D device

| | RTX 5080 | RTX 4090D |
|---|---|---|
| Architecture | Blackwell | Ada Lovelace |
| Boosted GPU clock / $Mhz$ | 2820 | 2700 |
| Boosted VRAM clock | $1875 Mhz$ GDDR7 | $1315 Mhz$ GDDR6X |
| Number of CUDA cores | 10752 | 14592 |
| Bus width / $bit$ | 256 | 384 |
| Bandwidth / $GB/s$ | 960.0 | 1008.4 |
| Number of transistors / million | 45600 | 76300 |
| FP16 / $TFLOPS$ | 1801 | 1321 |
| FP32 / $TFLOPS$ | 64.15 | 82.6 |